\documentclass[twocolumn,preprintnumbers,nopacs]{revtex4}
\usepackage{graphicx}
\usepackage{bm}
\usepackage{color}
\usepackage[normalem]{ulem}

\graphicspath{%
    {converted_graphics/}
    {/}
    }

\begin{document}

\title{Shortcut to adiabaticity in a Stern-Gerlach apparatus}

\author{Fran\c{c}ois Impens$^1$ and David Gu\'ery-Odelin$^2$}
\affiliation{$^1$ Instituto de F\'{i}sica, Universidade Federal do Rio de Janeiro,  Rio de Janeiro, RJ 21941-972, Brazil
\\
$^2$ Laboratoire Collisions, Agr\'egats, R\'eactivit\'e, IRSAMC, Universit\'e de Toulouse, CNRS, UPS, France}

\begin{abstract}
We show that the performances of a Stern-Gerlach apparatus can be improved by using a magnetic field profile for the atomic spin evolution designed through shortcut to adiabaticity technique.  Interestingly, it can be made more compact - for atomic beams propagating at a given velocity - and more resilient to a dispersion in velocity, in comparison with the results obtained with a standard uniform rotation of the magnetic field. Our results are obtained using a reverse engineering approach based on Lewis-Riesenfeld invariants. We discuss quantitatively the advantages offered by our configuration in terms of the resources involved and show that it drastically enhances the fidelity of the quantum state transfer achieved by the Stern-Gerlach device.
\end{abstract}

\maketitle

\section{Introduction}
\label{sec:introduction}

The Stern-Gerlach apparatus, used in the last century to obtain an experimental evidence of angular momentum quantization~\cite{Stern21}, has also interesting features for atom optics~\cite{Robert91}. In particular, this device has been used successfully in the first observation of a geometric phase~\cite{Berry84} in atom interferometry~\cite{Miniatura92PRL}. This system entangles the external atomic motion with the total angular atomic momentum, and produces a spatial separation between atomic wave-packets corresponding to different angular momenta. The Stern-Gerlach device can be used to perform transformations on the atomic spins, mapping initial angular momentum states to determined final states. 

These transformations are usually achieved with a magnetic field presenting an helicoidal profile~\cite{Miniatura92APB}. A simple way to map reliably initial spin states to well-defined target spin states is to design the magnetic field profile in such a way that the atomic spin follows the locally rotating magnetic field adiabatically in the course of its propagation through the device. The bottleneck of this approach is that
 the adiabatic regime imposes a minimum length over which the magnetic field may change its direction. This length is proportional to the atomic velocity and inversely proportional to the magnetic field modulus. Since strong magnetic fields may be undesirable, there is a trade-off between the speed of the adiabatic evolution and the magnetic field strength.

 The purpose of this article is to show that this trade-off can be greatly improved by using the shortcut to adiabaticity(STA) technique~\cite{STAReview13}. More precisely, we design a suitable magnetic field profile by using the reverse engineering approach based on the Lewis-Riesenfeld invariants~\cite{Lewis69,PRLSTA1_2010,NJP}. These methods enable one to guarantee a transitionless evolution faster than the time scale imposed by the adiabatic regime. The STA approach has been shown experimentally to efficently speed up the transport or manipulation of wave functions \cite{David08b,Bowler12,Walther12,Schaff10,Schaff11,Rohringer15,David13,David16} and even thermodynamical transformations \cite{Boltzmann,NaturePhys,APL}. Concerning the transfer of quantum states, recent impressive implementations have been reported in cold atoms experiments \cite{Bason12}, solid-state architectures~\cite{STASolidZhou17} or in opto-mechanical systems~\cite{STAOptoZhang17}. The transposition of those ideas to integrated optics devices has been recently explored \cite{Tseng12,Martinez14,Valle16}. 

Our proposal of STA-engineered Stern-Gerlach device outperforms the traditional rotating field in terms of speed of the quantum evolution for a given amount of resources - here the magnetic field -. The STA-engineered Stern-Gerlach is also robust toward  toward a dispersion in the atomic velocities, and may achieve very high fidelities in an atomic spin state transfer. We will illustrate our arguments by considering specifically the case of spin-one particles such as the hydrogen fragments issued from $H_2$ dissociation~\cite{Medina11,Medina12,Robert2013}.  This example is relevant, since an experiment based on an arrangement of Stern-Gerlach devices has been recently proposed to evaluate the spin-coherence of this dissociation~\cite{Carvalho15}. 
     
   The paper is organized as follows. In Section~\ref{sec:Standard Stern Gerlach}, we investigate the efficiency of a Stern-Gerlach device using a plain helicoidal configuration of the magnetic field. In Section~\ref{sec:Shortcut Adiabaticity}, we provide the general framework to determine a suitable magnetic field in a Stern-Gerlach device using STA technique. In Section~\ref{sec:examples STA}, we study an example of such a magnetic field profile suitable to realize fast and robust angular momentum evolution. In particular, we estimate the quantum fidelity and the speed-up of the spin transfer enabled by this configuration. We also study the resilience toward a dispersion in the atomic propagation times, and compare the performance of this configuration with respect to a standard Stern-Gerlach apparatus using an equivalent magnetic field.\\\

\section{Efficiency of a Stern-Gerlach device with an helicoidal magnetic field}
\label{sec:Standard Stern Gerlach}

In this Section, we review the propagation of a spin-one particle in a standard Stern-Gerlach device. This apparatus involves an helicoidal magnetic field which rotates of an angle $\pi/2$ over a certain length, corresponding to a propagation time $T$ for a given class of atomic velocities. An atom propagating at constant velocity along the helicoid axis experiences locally a uniform rotation of the magnetic field. We investigate the transfer of the angular momentum from the initial state $| J=1, m_z=1 \rangle$ to the target state $| J=1, m_x=1 \rangle$ as a function of the atomic propagation time $T$. 

For this purpose, the particle is subjected to the time-dependent Hamiltonian $\hat{H}(\mathbf{B}(t)) = - \gamma \: \mathbf{B}(t) \cdot \hat{\mathbf{J}}$ between the initial time $t=0$ and final time $t=T$. $\vec{\hat{J}}$ is the vectorial angular momentum operator of a spin-one particle and $\gamma$ accounts for the strength of the coupling and includes the atomic Land\'e factor. The magnetic field, $\mathbf{B}(t)$, in the comoving frame seen at the central atomic position~\cite{Footnote1}, reads $\mathbf{B}_{\rm st.}(t)=  \Re \left[ B_0 (\hat{\mathbf{z}}+ i  \hat{\mathbf{x}}) e^{i \frac {\pi} {2} \frac {t} {T}} \right] .$  As there is only a single characteristic time scale, namely the Larmor time $T_L=2 \pi  /  \gamma B_0 $ associated with the spin precession in the magnetic field, the efficiency of the quantum transfer may only depend on the ratio $T / T_L.$  We have used the Qutip package~\cite{Qutip1,Qutip2} to simulate the evolution of the initial spin state $| m_z=1 \rangle$ in a time-dependent rotating magnetic field. The efficiency of the spin transfer is quantified by the fidelity $\mathcal{F}$~\cite{Jozsa94} of the final state $| \psi( T )  \rangle$ to the target state $\mathcal{F}= | \langle m_x=1 | \psi( T )  \rangle |^2$. 

\begin{figure}[htbp]
\begin{center}
\includegraphics[width=8 cm]{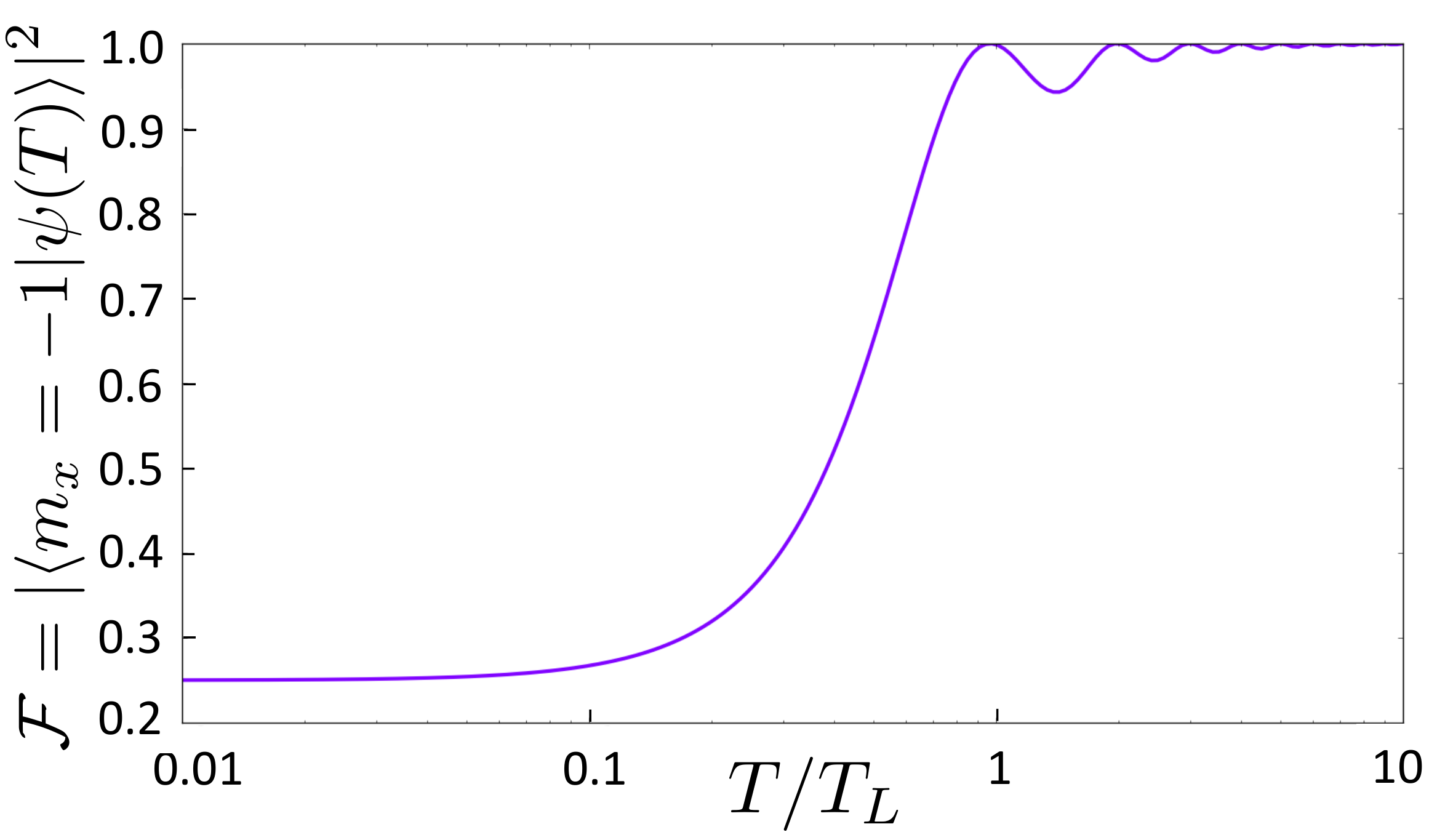}
\end{center}  \caption{Fidelity of the final state $\mathcal{F}= | \langle m_x=1 | \psi( T )  \rangle |^2$ as a function of the ratio $T/T_L$ between the total time and the Larmor time. This profile is independent of the considered Larmor pulsation, which has been set to $\omega_L \: = \:10^6 \: \: {\rm rad/s}$ in the numerical simulation.} 
\label{fig:FidelityStandardSternGerlach}
\end{figure}

The results are shown in Figure~\ref{fig:FidelityStandardSternGerlach}.  For short durations $T \ll T_L,$ the fidelity of the final state is barely higher than that of the initial state. As expected, in this limit, the atomic angular momentum state is almost unaffected by the Stern-Gerlach device: the rotation of the magnetic field is too fast to drive the atomic spin. On the other hand, for times $T \gg T_L,$ the atomic spin follows adiabatically the field, yielding a final state very close to the target state. A quantum fidelity equal to unity is achieved when $T=T_L.$ For total times $T$ {greater or equal} than a few Larmor times $T_L$, the fidelity stays higher than $99\%$. The Larmor time $T_L$, inversely proportional to the strength of the magnetic field, thus corresponds to a reliable spin transfer in a standard Stern-Gerlach apparatus. We shall use this duration to determine the figure of merit of a Stern-Gerlach device enhanced by the STA technique. Using typical experimental parameters for Stern-Gerlach atom interferometry with hydrogen fragments~\cite{Miniatura92PRL}, we consider atomic velocities $v \: \sim \: 10 \: \: {\rm km/s}$, a gyromagnetic ratio $\gamma \simeq \mu_b / \hbar$ and a magnetic field $B_0 \: = \: 0.1 \: {\rm G}$. a Larmor time
 $\nu_L \: = \: 0.14 \: {\rm MHz}.$ The corresponding Larmor time $T_L \: \simeq \:  7 \: \: {\rm \mu s}$ then yields the minimum length of a few $L\: \simeq \: 7 \: \: {\rm cm}$ for an efficient spin transfer in a Stern-Gerlach device with an helicoidal magnetic field. 

Finally, we note that the fidelity is a non-monotonic function of the ratio $T/T_L$. The small oscillations shown in Figure~~\ref{fig:FidelityStandardSternGerlach} are consistent with the predictions for the spin transition amplitudes of atomic spins experiencing a non-adiabatic evolution in an inhomogeneous magnetic field~\cite{Hight77}, evidenced experimentally in~for metastable hydrogen atoms~\cite{Hight78,Robert92}.

\section{Determination of the magnetic field from Lewis Riesenfeld invariants}
\label{sec:Shortcut Adiabaticity}

In this Section, we search for a magnetic field profile suitable to realize fast spin evolution of a spin-one particle in a Stern-Gerlach device, with an initial angular momentum along the $Oz$ axis and a target angular momentum state along the $Ox$ axis. We find the general equations for the field and determine suitable boundary conditions to be satisfied by the Lewis-Riesenfeld invariants~\cite{Lewis69}.

\subsection{General equations for the dynamical invariant}

For a given time-dependent Hamiltonian, $ \hat{H}(t)$, a dynamical invariant, $\hat I(t)$, fulfills the equation \cite{Lewis69}:
\begin{equation}
i \hbar \frac {\partial \hat{I}(t)} {\partial t} = [ \hat{H}(t),\hat{I}(t)].
\label{invariant}
\end{equation}
A natural choice for the rotation of a spin 1 is to search for a dynamical invariant in the form $\hat{I}(t) =  \mathbf{u}(t) \cdot \mathbf{J}(t)$ where the spin operators $\{ \hat{J}_{k}  \}$ with $k=x,y,z$ form a closed Lie algebra~\cite{LieAlgebra1,LieAlgebra2} as the Pauli matrices for SU(2) and $ \mathbf{u}(t)$ is a vector that needs to be determined. Interestingly, the dynamical operatorial equation (\ref{invariant}) can be translated in a simple linear differential equation 
describing a clockwise precession of the vector $\mathbf{u}(t)$ around the magnetic field $\mathbf{B}(t)$ 
\begin{equation}
\frac {d \mathbf{u}(t)} {dt} =  \gamma  \mathbf{u}(t) \times \mathbf{B}(t).
\label{precession}
\end{equation}
In the following,  we set the evolution of $\mathbf{u}(t)$ and infer from Eq.~(\ref{precession}) the explicit expression for $\mathbf{B}(t)$ \cite{DavidarXiv17}. This amounts to reverse engineer Eq.~(\ref{precession}).
Actually, we have a lot of freedom to choose the function $\mathbf{u}(t)$ {and this choice does not fully constrains the function $\mathbf{B}(t)$. In what follows, we fix $B_y=0$ and proceed to determine $\mathbf{B}(t)$ from Eq.~(\ref{precession})~\cite{Footnote2}. Finally, to connect the eigenstates at initial and final time of $\hat I (t)$ and $\hat H (t)$, we impose the following commutation relations $[ \hat{H}(0),\hat{I}(0)]=[ \hat{H}(T),\hat{I}(T)]=0$.  \\

\subsection{Resilience of the atomic spin transfer towards an atomic velocity dispersion}
\label{subsec:resilience}

Before proceeding, we highlight an interesting property arising from the commutation between the invariant and Hamiltonian operators at the final time, i.e. $[ \hat{H}(T),\hat{I}(T)]=0$. From the equation of motion~(\ref{invariant}), this condition implies that $\left. \frac {\partial \hat{I}}  {\partial t} \right|_{T} =0.$ Consequently, to leading order, 
 the eigenstates of the dynamical invariant $\hat{I}(T)$ are unaffected by a small fluctuation of the interaction time $T$ with respect to a reference value $T_0.$  By construction, the atomic spin is an eigenstate of the invariant operator at all times. One thus expects that the final atomic spin state should also be unchanged to leading order by a small fluctuation of the atomic time of flight. This property enables one to perform an efficient spin rotation over a broad range of particle velocities, turning the apparatus robust against an atomic velocity dispersion. In the next Section, we will verify numerically this resilience of the spin transfer for a concrete example of magnetic field.

\subsection{Relation between the magnetic field and dynamical invariant}

 The vector $\mathbf{u}(t)$ is conveniently parametrized in spherical coordinates as a unit vector, $\mathbf{u}(t)= (\sin \theta \cos \varphi, \sin \theta \sin \varphi, \cos \theta)$. From the precession equation (\ref{precession}), we get the components of the magnetic field as a function of the spherical angles $\theta(t)$ and $\varphi(t)$
 \begin{equation}
 \gamma B_x (t)  =  \frac {\dot{\theta}} {\sin \varphi}, \quad
  \gamma B_z (t) = -\dot{\varphi}  + \frac {\dot{\theta} \cos \theta \cos \varphi} {\sin \theta \sin \varphi} \, .  \label{eq:BfieldEqs} 
 \end{equation}
We shall thus search for acceptable angular functions avoiding divergences in the magnetic field. Typically, as seen from the equations above, such divergence may occur when the invariant pointer crosses the equatorial line defined by $\theta=0$ or the meridians defined by $\varphi=0,\pi$. This geometric constraint on the pointer trajectory sets a limit on the shortest spin transfer time achievable with the STA method when using polynomial angles of a certain degree. 

\subsection{Boundary conditions on the spherical coordinates of the invariant.}

We derive here the boundary conditions to be fulfilled by the angular functions $(\theta(t),\varphi(t)).$ These functions must enable the commutation between the invariant pointer and the Hamiltonian at the initial and final times. 
 Since the initial and target states must be eigenvectors of the initial and final Hamiltonian respectively, the magnetic field direction at these times is fixed according to  $\mathbf{B}(0)= B_z^I \: \hat{\mathbf{z}}$ and $\mathbf{B}(T)= B_x^F \hat{\mathbf{x}}$. The commutation between the invariant and Hamiltonian is then ensured by setting the pointer $\mathbf{u}(t)$ parallel to the magnetic field at these times. This is achieved by imposing the following conditions on the spherical coordinates
\begin{eqnarray}
\theta(0)=\pi, \quad \theta(T)=\pi/2, \nonumber \\ 
\varphi(0)=\pi/2, \quad \varphi(T)=0  \, . \label{eq:boundary1}
\end{eqnarray}
  We now search for suitable expansions of the angular functions near the initial and final times that yield through Eq.~(\ref{eq:BfieldEqs}) the magnetic field  $ B_x (0)=0$, $ B_z (0)=B_z^I$ and $B_x (T)=B_x^F$,  $B_z (T)=0.$ Using the perturbative expansion $\theta(t)=\pi+ \alpha \: t^m + o(t^m) $ and  $\varphi(t)=\pi/2+ \beta \: t^n + o(t^n) $ in the vicinity of $t=0$, one obtains that the condition on the initial magnetic field is equivalent to  $m>1$, $n=1$, and $\beta = - \gamma B_z^I / (m+n).$ The lowest-order expansion compatible with this condition corresponds to $n=1$ and $m=2$, that is to say 
\begin{equation}
\dot{\theta}(0)=0, \quad \ddot{\theta}(0)>0, \quad \dot{\varphi}(0)=  - \gamma B_z^I /3   \label{eq:boundary2} \,. 
\end{equation}

Using a similar expansion close to the final time with $\tau=t-T,$ $\theta(\tau)=\pi/2+ \alpha' \: \tau^{m'} + o(\tau^{m'}) $ and $\varphi(\tau)= \beta' \: \tau^{n'} + o(\tau^{n'}) $ with $\tau=T-t,$ one obtains that the second set of conditions is equivalent to $m' = n' + 1$,  $n'>1$ and $m' \alpha'=  \gamma \beta' B_x^F$. It can be fulfilled by choosing $n'=2$ and $m'=3$, yielding another set of conditions
\begin{eqnarray}
& \varphi(T)=0, &  \quad  \dot{\varphi}(T)=0, \nonumber \\
& \dot{\theta}(T)=0, &  \quad \ddot{\theta}(T)=0,  \quad \dddot{\theta}(T)= \gamma B_x^F \ddot{\varphi}(T)  \label{eq:boundary3}  \,.
\end{eqnarray}

\section{Example of fast magnetic spin driving with shortcut to adiabaticity.}
\label{sec:examples STA}

In this Section, we give an example of magnetic field profile realizing the STA, by finding suitable polynomial functions for the spherical coordinates of the invariant pointer. We show by numerical simulations that this magnetic field profile yields an extremely reliable transfer of a single quantum state.  We also discuss quantitatively the enhancement brought by the STA in a Stern-Gerlach device. For this purpose, one must consider the trade-off between the speed-up brought by the STA and the amount of resources involved~\cite{STAOptoZhang17}.  In contrast to the standard Stern-Gerlach device, the STA-engineered Stern-Gerlach involves a magnetic field with a time-dependent amplitude.  To work out explicitly the comparison between the two devices, we shall either consider as a resource the average magnetic field seen by the atom in the STA device, i.e. $B_{\rm av}= (1/T) \int_0^T dt ||\mathbf{B}(t)||,$ or the maximum magnetic field $B_{\rm max}= {\rm max} \{ ||\mathbf{B}(t)||,t \in [0,T] \}.$ These two criteria will provide different figures of merit.  As seen below, with both criteria the STA-engineered Stern-Gerlach outperforms the standard device.

\subsection{Example of suitable magnetic field profile.}

A simple way to match simultaneously the conditions~(\ref{eq:boundary1},\ref{eq:boundary2},\ref{eq:boundary3}) is to look for polynomials functions of the form $\theta(t)=P\left( \frac {t} {T} \right)$ and $\varphi(t)=Q \left( \frac {t} {T} \right)$. Lowest-order suitable polynomials can be obtained as:
\begin{eqnarray}
P(x) & = & \pi - 3 \pi x^2 + 4 \pi x^3 - \frac {3 \pi} {2}  x^4  \nonumber \\
Q(x) & = & \frac {\pi} {2} - \frac {\mathcal{B}_z^I } {3} x + \left(- 3 \pi - \frac {6 \pi} {\mathcal{B}_x^F}+ \mathcal{B}_z^I \right) x^2 \label{eq:polynomials}  \\
& + & \left( 4 \pi + \frac {12 \pi} {\mathcal{B}_x^F} -  \mathcal{B}_z^I  \right) x^3
+ \left(- \frac {3 \pi} {2}  - \frac {6 \pi} {\mathcal{B}_x^F} +\frac {\mathcal{B}_z^I } {3}  \right) x^4 \nonumber
\end{eqnarray}
where we have introduced the adimensional magnetic fields $\overrightarrow{\mathcal{B}}^I \: = \: \gamma \: T \: \mathbf{B}(0)$ and  $\overrightarrow{\mathcal{B}}^F \: = \: \gamma \: T \: \mathbf{B}(T).$ Let us stress that this choice of lowest-order polynomials is by no means unique, since one of the relations only constrains the ratio of the time derivatives of the angular functions. With this choice, the azimuthal angle $\theta(t)$ of the field is completely independent of the initial and final values of the magnetic field, which affect the longitude angle $\varphi(t)$.\\

At this stage, one can determine the full magnetic field profile from Eq.~(\ref{eq:BfieldEqs}). The values of the angle $\varphi(t)$ should be kept in the interval $]0,\pi[$ in order to avoid a divergence in the magnetic field~\cite{Footnote3}. For this purpose, one must choose carefully the sign of the magnetic field component $B_x^F$ at the final time.  Indeed, from Eq.(\ref{eq:polynomials}) one has $\dddot{\theta}(T)<0,$ so by virtue of Eq.(\ref{eq:boundary3}) one must have $B_x^F <0$  to ensure that $\varphi(T)=0$ is a local minimum. As a consequence of this choice, the considered device maps the initial state $| m_z=1\rangle$ to the target spin state $| m_x=-1\rangle$. The Figure below shows an example of magnetic field profile determined by Eq.(\ref{eq:polynomials}). For the considered parameters and with the propagation time $T \: = \: 2 \: {\mu s}$, an efficient state transfer is achieved in a STA-enhanced Stern-Gerlach with a maximum magnetic field $B_{\rm max} \:\simeq \: 0.22 \: {\rm G}$ to be compared with the magnetic field $B_{\rm st}^0= 2 \pi / \gamma T  \simeq \: 0.36 \: {\rm G}$ required in a standard Stern-Gerlach.
\begin{figure}[htbp]
\begin{center}
\includegraphics[width=8.8 cm]{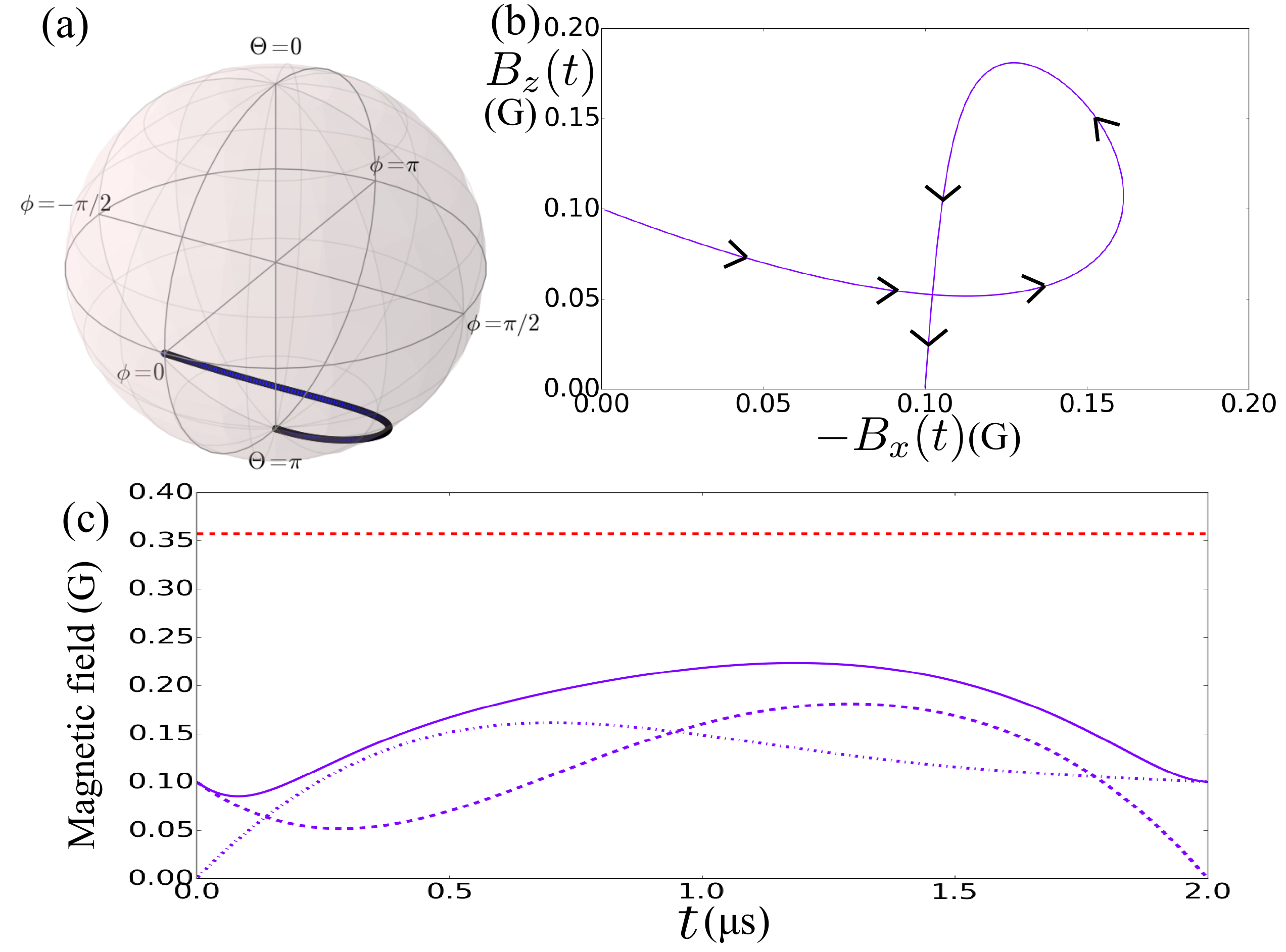}
\end{center}  \caption{Time-dependent magnetic field determined by reverse engineering with the Lewis-Riesenfeld invariant method for a total atomic propagation time $T \: = \: 2 \: {\mu s}$ and with the initial and final magnetic field components fixed to $B_z^I \: = \: 0.1{\rm G}$ and $B_x^F \: = \: - 0.1{\rm G}$ respectively. (a) Trajectory of the pointer vector $\mathbf{u}(t)$ on the unit sphere. The initial point is on the pole and ending and the final point on the equator. (b) Parametric plot of the magnetic field seen at the atomic central position. (c) Plot of the norm $|| \mathbf{B}(t) ||$ (solid purple line), of the $-B_x(t)$ (dash-dotted purple line) and $B_z(t)$ (dotted purple line) components of the magnetic field as a function of time. The dashed red line represents the larger magnetic field $B^0_{\rm st}$ enabling a reliable spin transfer in a standard device during the time $T$.}   
\label{fig:Magnetic Field Profile}
\end{figure}

\subsection{Fidelity and resilience of the spin transfer in a STA-engineered Stern-Gerlach device.}

Using the magnetic field defined in Eqs.~(\ref{eq:BfieldEqs}), we simulate~\cite{Qutip1,Qutip2} the temporal evolution of an atomic spin in a Stern-Gerlach device with a STA-designed magnetic field.  Precisely, we keep track of the expectation values of the angular momentum projections $\langle \hat{J}_z \rangle (t)$ and $\langle \hat{J}_x \rangle (t)$, as well of the fidelity of the atomic spin state with respect to the target state $ | m_x=-1 \rangle$. The latter may be written as $\mathcal{F}(t) = | \langle m_x=-1 | \psi(t) \rangle|^2$, where the quantum state $|\psi(t) \rangle=  \mathcal{T}  \exp \left[ - \frac {i} {\hbar} \int_0^T dt \hat{H}(\mathbf{B}(t)) \right]  | m_z=1 \rangle$ is the time-dependent atomic spin state evolved from the initial state $| m_z=1 \rangle$ through the interaction with the time-dependent magnetic field $\mathbf{B}(t).$ 

The corresponding results are shown on Figure~\ref{fig:SpinExpectation} and Figure~\ref{fig:FidelitySTASternGerlach} respectively. We first consider a STA Stern-Gerlach device and a standard Stern-Gerlach apparatus using equivalent magnetic fields. When the atomic spin propagates in a standard Stern-Gerlach device, whose magnetic field is $B_{\rm av}= (1/T) \int_0^T dt ||\mathbf{B}(t)||,$ the expectation value $\langle \hat{J}_x \rangle (T)> -1$ and the relatively low fidelity achieved ($\mathcal{F} \simeq 0.60$) reveal an imperfect transfer to the target state $ | m_x=-1 \rangle$. Even when the maximum field modulus $B_{\rm max}$ is used in the standard Stern-Gerlach device, the fidelity of the spin state to the target state saturates at the value $\mathcal{F} \simeq 0.80$.  In contrast, when the atomic spin propagates in the STA-designed magnetic field, the atomic spin projections expectation values are very close to $\langle \hat{J}_z \rangle (T)=0$ and $\langle \hat{J}_x \rangle (T)  =-1,$ showing that the final spin state is very close to the target state $| m_x=-1 \rangle$ at the final time $T$. With a standard Stern-Gerlach device using a larger magnetic field $B_{\rm st}^0$, one may achieve a good fidelity in the quantum state transfer over a large interval of propagation times $t$.

Nevertheless, the behaviour of the fidelity near the optimal time is different for the standard and for the STA Stern-Gerlach, enabling the latter to reach very high fidelities in a single state transfer.  Indeed, the error committed in this transfer $\epsilon(t)=1-\mathcal{F}(t)$ decreases sharply for the STA device and can become arbitrarily low in the vicinity of the ideal propagation time $T$. Differently, for the standard SG device a finite error remains even at this ideal time. Its value depends on the magnetic field involved. In order to obtain a quantitative measure of the reliability of the spin transfer with a STA device, we analyse different fidelity thresholds for propagation times $t$ close to the ideal time $T$. Noting $\Delta t= t-T$ the time propagation mismatch, the inset of Figure~\ref{fig:FidelitySTASternGerlach} reveals that the error committed can be as low as $\epsilon(t) \leq 10^{-8}$ for  $ | \Delta t / T | \leq 2 \times 10^{-3} $, or $\epsilon(t) \leq 10^{-5}$ for  $ | \Delta t / T | \leq 1.5 \: \times 10^{-2},$ and $\epsilon(t) \leq 10^{-2}$ for $  |\Delta t / T| \: \leq 0.1 .$

Equivalently, these fidelities can be achieved for a certain velocity interval $\Delta v=v-v_0$  around a reference velocity $v_0$ for which the STA magnetic field has been designed. The resilience of the STA and standard Stern-Gerlach devices increases with the strength of the magnetic field involved. With a maximum magnetic field of $B_{\rm max} \: = \: 0.22 \: {\rm G}$, the STA Stern-Gerlach achieves a fidelity $\mathcal{F} \geq 99 \%$ for a class of velocities $\Delta v$ such that $| \Delta v  /v_0| \leq 10 \%$. This corresponds to a velocity spread of $\Delta v \simeq 1 {\rm km/s}$ for the experiments~\cite{Hight78,Miniatura92APB} or for the slow beams of metastable hydrogen obtained from molecular dissociation in~\cite{Medina12,Robert2013}, or $\Delta v \simeq 4 {\rm km/s}$  for the fast hydrogen beams~\cite{Medina11,Robert2013}.

These measures can be compared with the reliability threshold for a universal set of quantum gates in order to obtain scalable quantum error correction~\cite{Shor96,Knill98,Preskill98}. As shown in these references, the availability of a finite set of gates enabling universal quantum computation and with a probability of failure below a certain threshold indeed enables one to implement large scale quantum error correction. The exact value of this threshold depends on a variety of factors such as the structure of the code employed and on the noise model~\cite{Gottesman09}. Earlier estimates of this threshold gave a maximum failure probability of $p \simeq 2.10^{-5}$~\cite{Preskill06} for seven-qubit codes. Using a different topology, surface code quantum computing~\cite{Fowler12,MartinisNature14} enables efficient quantum computation with gates fidelities of $\mathcal{F} \geq 99 \%$. 

The Stern Gerlach device enhanced by the STA technique may thus obtain the transfer of a single quantum state with a reliability over this threshold for a significant range of propagation times, suggesting that this apparatus could be used within a quantum computing architecture.

\begin{figure}[htbp]
\begin{center}
\includegraphics[width=8 cm]{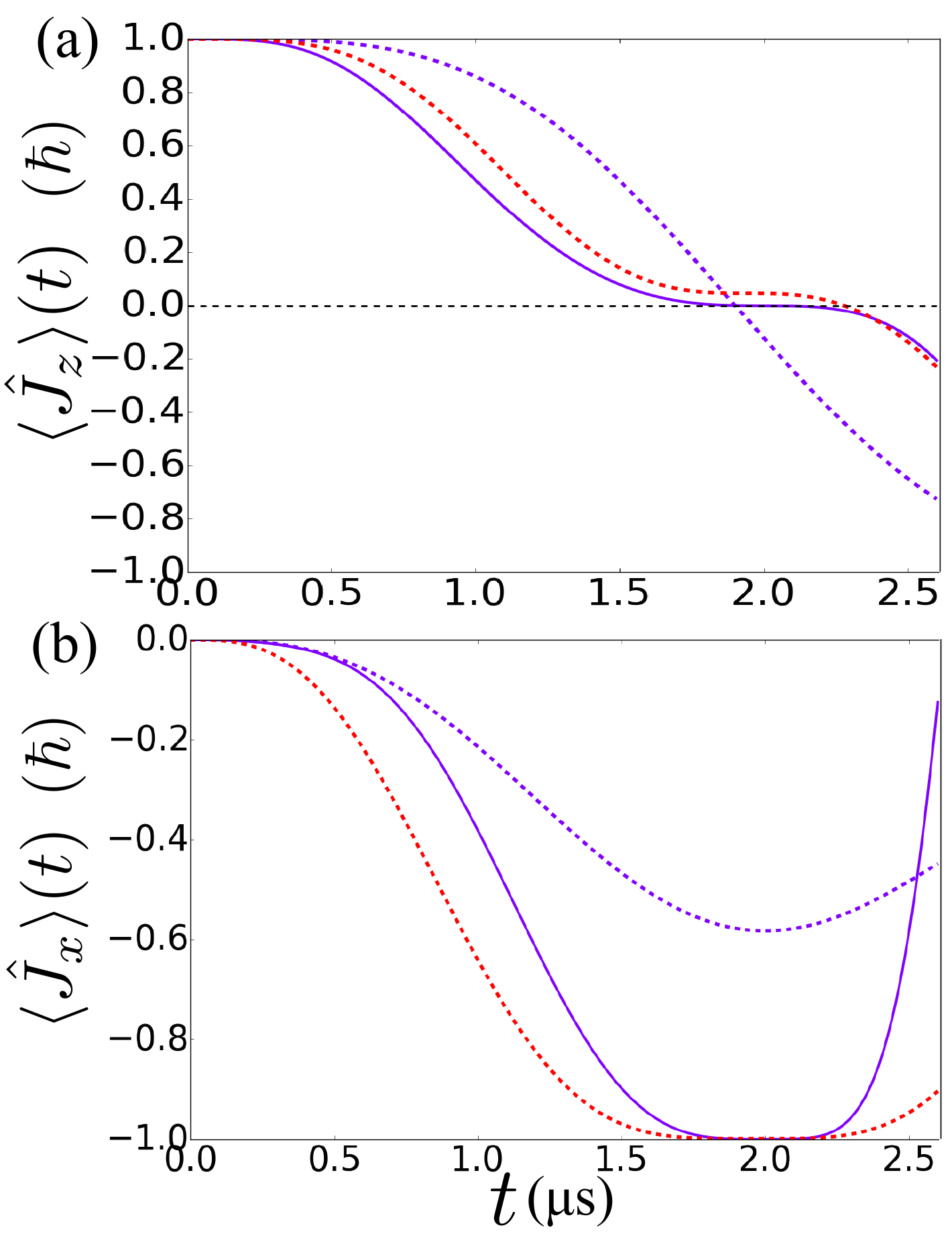}
\end{center}  \caption{ Average values of the angular momentum projections (a) $\langle \hat{J}_z \rangle(t)$ and (b) $\langle \hat{J}_x \rangle(t)$ in units of $\hbar$ as a function of the propagation time $t$. We consider the propagation in a STA-enhanced Stern-Gerlach device (solid purple lines), in standard Stern-Gerlach devices using either a magnetic field equal to the temporal average of the STA magnetic field (dashed purple lines) either a larger magnetic field $B^0_{\rm st}$ enabling an efficient spin transfer during the time $T \: = \: 2 \: {\mu s}$ (dashed red lines). Purple lines represent the use of equivalent resources in terms of magnetic field. The numerical parameters and magnetic fields are identical to those used in Figure~\ref{fig:Magnetic Field Profile}.} \label{fig:SpinExpectation} 
\end{figure}

\begin{figure}[htbp]
\begin{center}
\includegraphics[width=9 cm]{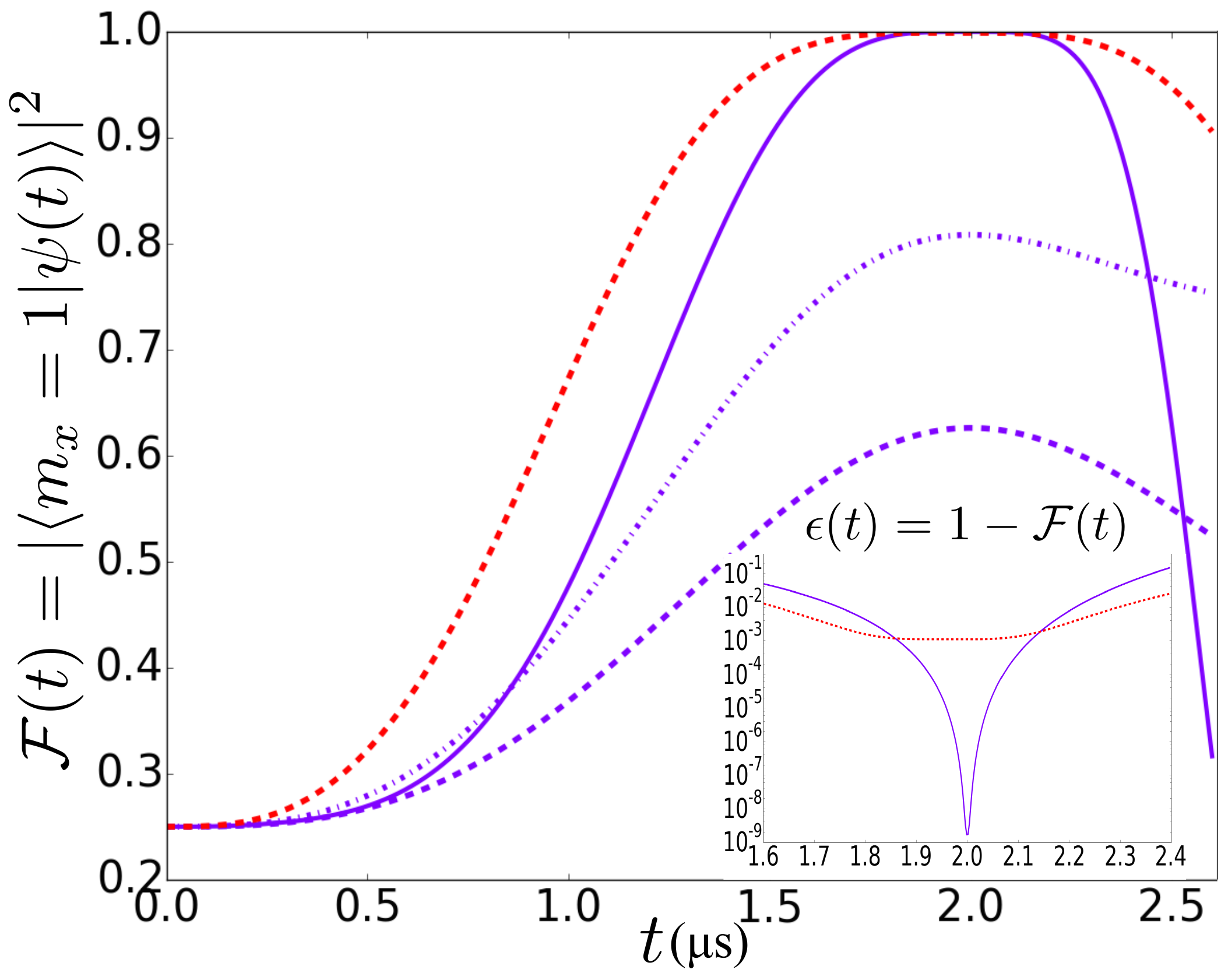}
\end{center}  \caption{Fidelity of the atomic spin state during its propagation in the Stern-Gerlach device: STA device (solid purple line), in the standard Stern-Gerlach apparatus with a magnetic field of modulus $B_{\rm av}$ (dotted purple line), $B_{\rm max}$ (dash-dotted purple line), and $B_{\rm st}^0$ (dotted red line). Purple lines represent the use of equivalent resources in terms of magnetic field, while the dotted red line corresponds to a stronger magnetic field $B_{\rm st}^0.$  The inset gives a logarithmic plot of the error committed $\epsilon(t)=1-\mathcal{F}(t)$ in the quantum state transfer for times $t$ in the vicinity of the reference time $T= 2 \:  {\mu s}$. The magnetic fields correspond to the parameters of Figure~\ref{fig:Magnetic Field Profile}.} 
\label{fig:FidelitySTASternGerlach}
\end{figure}

\subsection{Speed-up offered by the STA-engineered magnetic field.}

In order to evaluate quantitatively the benefits offered by the STA-engineered magnetic field over the standard helicoidal configuration, we estimate the resources -- in terms of magnetic field -- required to achieve a perfect spin transfer in a given total propagation time. As  previously, we consider either the average or the maximum magnetic fields involved in the STA configuration to determine the equivalent magnetic field in the standard STA device. For a range of times $T$, we derive the STA-engineered magnetic field from Eq.(\ref{eq:polynomials}), which gives readily the average and maximum magnetic fields involved. As seen previously, the magnetic field required in the standard device is inversely proportional to the propagation time $T$.

Figure \ref{fig:EfficiencySTAResources} compares the performances of the STA-engineered and the standard Stern-Gerlach apparatus from two equivalent points of views. Figure~\ref{fig:EfficiencySTAResources}(a) shows magnetic field required in both devices for a range of atomic spin transfer time $T$, while Figure~\ref{fig:EfficiencySTAResources}(b) reveals the speed-up offered by a STA Stern Gerlach in comparison with a standard Stern Gerlach loaded with a magnetic field of similar strength. These figures reveal that for a total transfer time $T  \gtrsim 0.8 \: {\rm \mu s},$ the STA-engineered Stern Gerlach device is more performant. The maximum magnetic field involved in the STA device is smaller than the magnetic field required in a standard Stern-Gerlach device. Equivalently, the standard Stern-Gerlach requires a longer atomic propagation time when using a magnetic field of modulus equal to the maximum STA-magnetic field. Note that the curves associated with the maximum magnetic field in the STA device present a slope discontinuity~\cite{Footnote4}.

On the other hand, when the atomic spin transfer time is such that  $T \lesssim 0.8 \: {\rm \mu s},$ the standard Stern-Gerlach becomes more efficient. Indeed, close to $T=0.5 \: {\rm \mu s}$, the magnetic field involved in the STA device diverges. This issue is related to the divergences generated by the roots of the angular functions $\theta(t)$ and $\phi(t)$, which appear in the interval $]0,T[$ when $T$ goes below a certain value. By using polynomial functions of higher order, it is possible to go to shorter times, while preserving a small enough magnetic field. When considering a family of polynomials of a given order, these divergences set a lower bound on the times achievable by the STA-engineered Stern Gerlach devices. 

\begin{figure}[htbp]
\begin{center}
\includegraphics[width=7.5 cm]{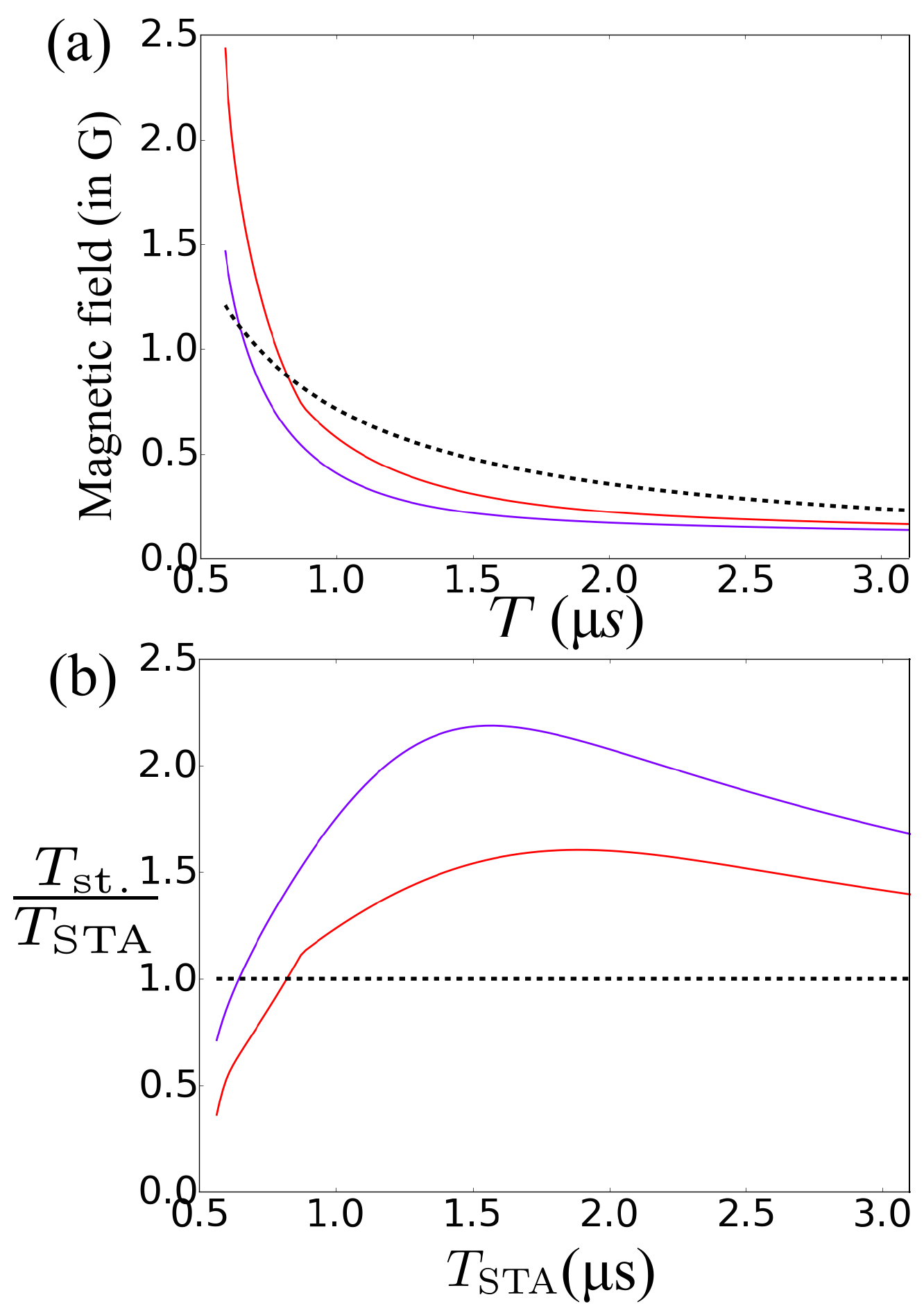}
\end{center}  \caption{   (a) Maximum (solid red line) and average (solid purple line) magnetic fields used in a STA-engineered Stern-Gerlach device, compared to the magnetic field in a standard Stern-Gerlach device (dotted black line), in order to perform an efficient spin transfer during the total time $T$.  (b) Speed-up offered by the STA configuration. Ratio between the transfer time $T_{\rm st.}$ required in a standard Stern-Gerlach device to the time $T_{\rm STA}$ required in a STA device. The standard Stern-Gerlach uses either a magnetic field of modulus $B_{\rm max}$ corresponding to the maximum field in the STA device (red line) or a magnetic field of modulus $B_{\rm av}$ corresponding to the average field in the STA device (purple line).  } 
\label{fig:EfficiencySTAResources}
\end{figure} 
 

To conclude, we have proposed to enhance the performances of a Stern-Gerlach device by using the technique of shortcut to adiabaticity. We have considered the propagation of spin-one particles in the device. Using the Lewis-Riesenfeld invariant approach, we have found a suitable magnetic field by reverse-engineering of the dynamical equation of motion for the invariant. The commutation between the invariant pointer and the Hamiltonian at initial and final times reduces the sensibility of the final state to the total propagation time. Using numerical simulations, we have demonstrated the validity of our approach, and provided a quantitative picture of the enhancement offered by this technique. The STA-engineered Stern-Gerlach apparatus appears to offer a better trade-off between time of propagation and magnetic field involved in the device. This conclusion is generic and valid for higher angular momentum. It provides the physical limits for the miniaturization of Stern and Gerlach device to entangle external and internal degrees of freedom. Finally, the STA-engineered Stern-Gerlach apparatus may achieve for the transfer of a single atomic spin extremely high fidelities for a narrow range of velocities, and fidelities above $99\%$ over a broad range of velocities. Such fidelities are below the error threshold for scalable quantum computation with surface codes. The fidelity enhancement provided by the STA method may open the possibility to use the Stern-Gerlach devices in the context of quantum information processing.

\acknowledgments

F.I. acknowledges a very fruitful collaboration and enlightening discussions on atom optics and Stern-Gerlach interferometry with Carlos Renato de Carvalho, Ginette Jalbert and 
Nelson Velho de Castro Faria.

\bibliography{170913biblio_geral}{}

\begin{thebibliography}{46}
\bibitem{Footnote1} We consider here the non-relativistic regime and neglect the variation of the magnetic field over the extension of the atomic wave-packet.
\bibitem{Footnote2} Note that one may construct a physical static magnetic field $\mathbf{B}(\mathbf{r})$ such that the function $\mathbf{B}(t)$ corresponds to the magnetic field seen by the atom in the non relativistic limit, i.e. $\mathbf{B}(t) = \mathbf{B}(\mathbf{r}(t))$ with $\mathbf{r}(t)$ the particle position. Indeed, Eq.(\ref{precession}) only constrains the gradient of the magnetic field along the direction of the atomic motion. Other components of the magnetic field gradient may be chosen at will to ensure consistency with Maxwell equations.
\bibitem{Footnote3} At the final time the singularity arising from the cancellation $\varphi(T)=0$ is cured by a simultaneous cancellation of the derivative $\dot{\theta}(T)=0$.
\bibitem{Footnote4}  A slope discontinuity occurs in Figure~\ref{fig:EfficiencySTAResources} because the maximum magnetic field, namely $B_{\rm max}(T)= \max \{ ||\mathbf{B}(t,T)||  \: \: | \: \: t \in ]0,T[ \}$, is not a continuously differentiable function of $T$. The STA magnetic field, obtained through a reverse engineering procedure depending on the total duration $T$, can be seen as a function of two variables $\mathbf{B}(t,T)$. Close to the singularity $T_S \:  \simeq \: 0.8 \: {\rm \mu s}$, the function $t \rightarrow ||\mathbf{B}(t,T)||$ has two distinct local maxima in the interval $]0,T[$, associated to the instants $t_1(T)$ and $t_2(T)$ such that  $t_1(T) < t_2(T)$.  The profile of this function depends smoothly on the total time $T$. For the specific value $T=T_S$ these two local maxima are equal, i.e. $||\mathbf{B}(t_1(T_S),T_S)||=||\mathbf{B}(t_2(T_S),T_S)||$. For $T<T_S$ one has $||\mathbf{B}(t_1(T),T)||>||\mathbf{B}(t_2(T),T)||$ and thus $B_{\rm max}(T)=||\mathbf{B}(t_1(T),T)||$. For $T>T_S$, one has $B_{\rm max}(T)=||\mathbf{B}(t_2(T),T)||.$ Since the difference $|t_1(T_S)-t_2(T_S)|>0$ is finite, one obtains a slope discontinuity for $B_{\rm max}(T)$ in $T=T_s$. This also induces a slope discontinuity in the ratio $T_{\rm st.}/T_{\rm STA}$ associated to a standard Stern-Gerlach device using a field of modulus $B_{\rm max}(T_{\rm STA})$.
\end{thebibliography}

\end{document}